# *Network Dynamics of Emotional Processing: A Structural Balance Theory Approach*

Sepehr Gourabi[1], Parinaz Khosravani[2], Shahrzad Nosrat[3], Roya Mohammadi[4], Masoud Lotfalipour[5]

1 Department of Psychology, Binghamton University, 4400 Vestal Parkway East, Binghamton, NY 13902, USA

2 Department of Psychology, Binaloud Institute of Higher Education, Mashhad, Iran

3 Student Research Committee, Kashan University of Medical Sciences, Kashan, Iran

4 Department of Mathematics and Computer Science, Amirkabir University of Technology, Tehran, Iran

5 Institute for Cognitive and Brain Sciences, Shahid Beheshti University, Tehran, Iran

**ABSTRACT**

Understanding emotional processing in the human brain requires examining the complex interactions between different brain regions. While previous studies have identified specific regions involved in emotion processing, a holistic network approach may provide deeper insights. We use Structural Balance Theory (SBT) to investigate the stability and triadic structures of signed brain networks during resting-state and emotional processing, specifically in response to fear-related stimuli. We hypothesized that imbalanced triadic interactions would be more prevalent during emotional processing, especially in response to fear-related stimuli, potentially reflecting the brain's adaptation to emotional challenges. We analyzed fMRI data from 138 healthy, right-handed participants and extracted functional connectivity matrices to assess positive and negative links, balanced and imbalanced triads, and the tendency to make hubs (TMH). We found that emotional processing was marked by an increase in positive connections and a decrease in negative connections compared to the resting state. Our findings clearly show that balanced triads significantly decreased, while imbalanced triads increased, indicating a shift towards instability in the brain's functional network during emotional processing. Additionally, the number of influential hubs was significantly lower during fear processing than in neutral conditions, suggesting a more centralized network and higher levels of network energy. These findings reveal the brain's remarkable adaptive capacity during emotional processing, demonstrating how network stability dynamically shifts through changes in balanced and imbalanced triads, hub tendencies, and energy dynamics. Our research illuminates a complex mechanism by which the brain flexibly reconfigures its functional network in response to emotional stimuli, particularly under fear conditions, with potential implications for understanding emotional resilience and neurological disorders.

# 1. INTRODUCTION

Understanding the intricate dynamics of emotional processing in the human brain remains a fundamental challenge in cognitive science. However, a comprehensive understanding necessitates adopting a more holistic perspective of the brain as a complex network with distributed processing [1–3]. Emotions have played a significant role in the process of evolution. Numerous theories have been proposed to explain the nature of emotions [4–6]. From an evolutionary standpoint, emotions facilitate the synchronization of various processes to address pressing and immediate concerns [7]. While early research emphasized the importance of the hypothalamus and its projections in emotional behavior [8], the Amygdala has been identified as a crucial component in responding to threatening and fearful situations [9].

For example, studies in the field of processing the emotion of fear indicate that the limbic circuit has been conformed with this part of the human brain regarding emotions [10]. Previous studies have indicated that fear is the most commonly expressed emotion in response to immediate and severe threats [11]. Fear is distinguished by a heightened motivational intensity, leading to a powerful inclination to evade or retreat from a particular stimulus [12]. Several theories suggest that fear is an innate or learned emotion in all humans and many other animals [13]. Furthermore, some of these theories propose different categories of fear. One argument posits that distinct neural systems are responsible for fear of pain, fear of predators, and fear of aggressive individuals within the same species [14].

This means that in most studies, researchers have attempted either to explain the cause of behavior based on a specific type of emotion or to elucidate the brain mechanism behind it. Sometimes, they have tried to identify the brain regions associated with emotion-processing. For example, in a study conducted using the HCP dataset and the emotion-processing task, regions of interest such as the Amygdala, hippocampus, insula, and medial prefrontal cortex were found [15]. However, are only these limited brain regions responsible for processing emotions?

An fMRI study that has used a machine learning method in order to investigate subjective experience of fear has indicated Increased activation in regions including the Amygdala and its surrounding areas, as well as the anterior insula, anterior midcingulate cortex (aMCC), thalamus, periaqueductal gray (PAG), midbrain, ventrolateral prefrontal cortex, lateral orbitofrontal cortex, and fusiform/ventral occipital-temporal regions [16]. Studies employing transcranial magnetic stimulation have revealed that fear diminishes when the dorsolateral PFC is stimulated [17]. Furthermore, In Social Anxiety Disorder (SAD), studies have noted reductions in Gray Matter Volume (GMV) in the right thalamus, the left parahippocampus, and the bilateral putamen. Conversely, a different pattern has been observed with larger subcortical GMVs and smaller cortical GMVs [18]. Specifically, individuals with Generalized Anxiety Disorder (GAD) show an increase in the right putamen volume along with decreases in the lateral/medial Prefrontal Cortex (PFC) and the left insula volumes when compared to anxiety disorders primarily characterized by fear [17].

To properly understand the complexity of the brain network, we must consider the interconnected links simultaneously, in line with the emergent property of complex systems like the brain. This principle highlights that components of a complex system do not function in isolation but are part of a greater whole, where the whole system's behavior is more than the sum of its parts. For this reason, we decided to approach a holistic perspective on the brain's processing of emotions using the structural balance theory. This approach aims to enhance our understanding of overall changes in the brain during the processing of emotions.

Based on triadic associations, the Structural Balance theory (SBT) provides a valuable framework for examining network properties [19,20]. Triadic associations refer to the relationships between three nodes in a network. SBT posits that a third node in a triad structure can significantly impact the relationships between the other nodes [21]. The theory suggests that the relationship between any two nodes is likely to align with the overall sign of the triad. Balanced triads (+, +, +) or (+, -, -) are more likely to result in positive relationships between nodes, whereas other triad configurations tend to lead to negative relationships [22].

The significance and applicability of SBT extend across various domains, including social [23,24], Psychological [25], political [26], and biological networks [27]. In recent years, SBT has been applied in the field of neuroscience. Researchers have utilized SBT concepts to investigate brain functional networks, particularly comparing networks in Healthy adults to those in disordered conditions such as obsessive-compulsive disorder (OCD) [28] and autism spectrum disorder (ASD) [19]. SBT allows for studying balanced and unbalanced triads and their topology to determine the stability of brain networks [20,21].

In the latest study conducted with this approach, only negative connections in the brain during emotional processing were examined using the analysis of EEG signals [29]. Now, our goal is to investigate additional components in the overall changes in the brain during this processing. By examining the interplay of balanced and unbalanced triads in brain networks, the study of SBT offers a nuanced understanding of how the brain adapts to different emotional states. This sheds light on the neural dynamics underlying emotional processing, which could have significant implications for understanding and potentially treating emotional disorders and enhancing emotional intelligence and resilience.

In this study, we employ the principles of SBT to examine the change in stability and triadic structures of signed brain networks in both resting-state and emotion-processing states. We hypothesize that imbalanced triadic interactions, which refer to triads with a mix of positive and negative relationships, will be prevalent during emotional processing, especially in response to fear-related stimuli. This imbalance may heighten emotional responses, impacting the brain's adaptability and potentially influencing human behavior.

### 1.1. Structural Balance Theory

SBT is a framework that originated from the work of Fritz Heider in the 1940s, aimed at comprehending the dynamics of interpersonal relationships and social structures [30,31]. This theory was formalized mathematically by Cartwright and Harary, making it a fundamental concept in network science [32]. At its core, SBT delves into networks where links can be either positive (e.g., signifying friendship or cooperation) or negative (e.g., denoting hostility or conflict) [33]. These networks, often called "signed networks," are characterized by positive and negative connections among entities [22,34]. This duality of links is a crucial feature of SBT as it examines the intricate interplay between these positive and negative connections within networks.

### 1.2. Type of Triads

SBT emphasizes triads, groups of three nodes (entities), and their connections. Within this framework, four distinct types of triads are recognized [22]:

**Strongly Balanced [+ + +]:** These triads consist of three positive links, indicating that all three entities have harmonious relationships. **Weakly Balanced [+ − −]:** These triads have two negative and one positive links, resulting in a subtle balance between conflict and cooperation.
**Strongly Unbalanced [+ + −]:** Triads of this type contain two positive and one negative link, representing a significant imbalance in the network. **Weakly Unbalanced [− − −]:** These triads comprise three negative links, signifying that all three entities are involved in a conflict.

### 1.3. Balance-Energy

In SBT, a network is deemed balanced when all its triads are strongly or weakly balanced. In contrast, an unbalanced network exhibits an overall tension and tends to evolve towards achieving a more balanced configuration.

To quantify the degree of balance within a network based on its triads, SBT introduces the concept of "balance-energy," initially proposed by Marvel et al. Balance-energy is a quantitative measure of balance within a network [35]. According to the minimum energy principle, higher balance-energy indicates more unbalanced triads [21]. As Marvel and colleagues proposed, Dividing the sum of negative ternary interactions by the total number of interactions gives the network's total energy, U [35]. Explicitly,

$$U = -\frac{1}{\binom{N}{3}} \sum_{ijk} S_{ij} S_{ik} S_{jk}$$

Where the summation is performed on all possible triads within the network, the connection between node i and node j is denoted by $S_{ij}$. $S_{ij}\ S_{ik}\ S_{jk}$ represents the multiplication of edge values in a triad. This product results in the balanced energy of the triad. The balance-energy equals -1 and +1 for balanced and imbalanced triads. The minus sign in front of the multiplication helps to provide a better understanding of the equation from a physical energy perspective. The term $\binom{N}{3}$ represents the number of 3 combinations from N elements, equivalent to the number of possible triads in an N-node network. This term plays a normalization role and confines the balance-energy between -1 and +1.

### 1.4. Tendency to Make Hub

The concept of "Tendency to Make Hub" (TMH) is a novel measure of global hubness, which Saberi and colleagues initially introduced to quantify the tendency of nodes in a network to become hubs (Saberi et al., 2021a). The TMH measure assesses the network's tendency to form hubs through positive and negative connections. Higher TMH values indicate a greater tendency for links to cluster around nodes and create hubs. TMH of the networks is defined as follows:

$$\text{TMH} = \frac{\sum_{i=1}^{N} D_i^2}{\sum_{i=1}^{N} D_i}$$

$D_i$ represents each node's degree, and $N$ represents the total number of nodes in the network.

## 2. METHOD

### 2.1. Functional Connectivity

Initially, we needed to extract the functional connectivity matrix for each subject in two states, resting-state and emotion-processing task, to examine our hypotheses through structural balance theory.

For the resting-state data, we initially extracted and loaded time series for each participant. The process continued based on these time series, representing changes in BOLD signals over time in this temporal framework. Then, Pearson correlation coefficients were calculated using the Glasser 360 atlas and mapping the 360-defined points in this atlas with the data of each individual. These correlation coefficients measure the strength and direction of linear relationships between different variable pairs. In functional connectivity, higher correlation values indicate more robust functional connectivity or more significant similarity in the activity patterns of related brain regions. The obtained matrix for the resting-state was diagonal and symmetrical.

However, participants responded to a task for emotion-processing data, so the connectivity matrix had to be extracted based on that. Initially, the time series and temporal information related to the task presentation and temporal features of the stimuli presented during the response were extracted. The process begins by loading the time series data and information related to events corresponding to the task and specified conditions. Time series data reflects aspects of brain activity at specific time points. In succession, event information and details about experimental events, including temporal features and classification of presented stimuli or conditions during the task, are included.

After acquiring the relevant data, the time series information undergoes a classification process into predicted discrete executions on the desired frames specified and derived from the event information. This classification is performed to separate data related to individual task executions and facilitates focused analysis.

When data is divided into discrete executions, the time series data continues to be appended across executions and merged along the final axis. This merging results in a composite array, namely "task_data," which provides a continuous time series encompassing the task and specified conditions.

The final step involves calculating the connectivity matrix. This function calculates Pearson correlation coefficients between variable pairs within "task_data." The resulting matrix serves as a connectivity image of the task between various brain regions or nodes in the temporal domain. Each element within the matrix conveys the strength and direction of correlation between corresponding variable pairs, potentially indicating interactions in different brain measurements.

## 2.2. Statistical Analysis

After extracting functional Connectivity matrices and the number of positive and negative links, we utilized the Interquartile Range (IQR) technique to remove any possible outliers in all examined parameters. The group-level normality of the measures was tested using the Shapiro-Wilk normality test. Then, we used the nonparametric Wilcoxon signed-rank test to statistically compare each paired group due to the non-normal distribution of the data. Subsequently, we conducted the pairwise analysis to compare the number of positive and negative links between the resting, fearful, and neutral states, followed by a thorough examination of the balanced and imbalanced triads present within the networks, as well as the TMH and the Structural balance-energy exhibited by the brain network in its respective states. A threshold of $p < 0.05$ (Fisher permutation) was used to determine statistically significant differences among the three groups. Significance levels were categorized into different groups based on the P value, and for each level, we indicated the number of stars in the boxplot chart. If p-value < 0.001, we denoted it with ***, if p-value < 0.01, with **, and if p-value < 0.05, with *.

## 2.3. Participants

To explore our hypothesis and apply the proposed theory, we utilized the Human Connectome Project (HCP) dataset, explicitly focusing on Functional magnetic resonance imaging (fMRI) data from 138 out of 1200 subjects. Due to the significant volume of HCP data and our limited analysis capabilities, we leveraged cloud-hosted data, facilitating our work. The data, made available by the Neuromatch Academy—specializing in computational neuroscience studies—was accessible in the Google Colab environment, covering 339 out of the total 1200 subjects. We refined the dataset to include only healthy, right-handed male participants. Using the Edinburgh Handedness Questionnaire to determine handedness [36], we selected exclusively right-handed individuals. In the final phase, participants' mental health was assessed using the Mini-Mental State Examination (MMSE), excluding those with scores below 24, considered outside the normal and healthy range [37]. This process resulted in a final cohort of 138 healthy adult subjects aged 22-35.

## 2.4. Data collection and preprocessing

### 2.4.1. Task

This task, adapted from the work of Hariri and colleagues [38,39], serves as the foundation for our study. Participants are presented with blocks of trials in which they must decide which of the two faces or shapes presented at the bottom of the screen matches the top image. This adaptation allows us to build on previous research and explore new avenues in our study.

The faces have either an angry or fearful expression. Trials are presented in blocks of 6 trials of the same task (face or shape), with the stimulus presented for 2000 ms and a 1000 ms inter-trial interval (ITI). Each block is preceded by a 3000 ms task cue ("shape" or "face") so that each block lasts 21 seconds, including the cue.

Each of the two runs contains three face blocks and three shape blocks, with 8 seconds of fixation at the end of each run. A bug was written into the E-prime script for the EMOTION task, so the task stopped before the last three trials of the last task block in each run. The bug in data collection was noticed when several participants had already collected data. As a result, the BOLD plots and E-Prime data for the EMOTION task are shorter in duration compared to what was originally intended in our design.

### 2.4.2. Resting-state

Four rfMRI data runs were collected, each lasting approximately 15 minutes. Two runs were conducted in one session, while the other two were conducted separately. Participants were instructed to keep their eyes open and fixate on a bright cross-hair projected onto a dark background in a darkened room to minimize visual distractions. Within each session, the oblique axial acquisitions alternated between right-to-left (RL) and left-to-right (LR) phase encoding directions. This data acquisition protocol allowed us to examine brain activity and connectivity patterns during relaxed fixation, providing valuable insights into the resting-state brain.

### 2.4.3. Preprocessing

The HCP preprocessing pipelines received an upgrade in Version 3, which involved implementing various software changes. Specifically, FSL, FreeSurfer, and Connectome Workbench were upgraded to versions 5.0.6, 5.3.0-HCP, and 1.1.1, respectively. These software updates have led to observable enhancements in the outputs generated by the HCP preprocessing pipelines.

We utilized the HCP dataset which underwent throughout this pipeline (WU-Minn, 2017): The data underwent several preprocessing steps to ensure accurate and high-quality results. These steps included gradient distortion correction, FLIRT-based motion correction using the "SBRef" volume as the target, TOPUP-based field map preprocessing using Spin echo field map for each BOLD run, distortion correction, and EPI to T1w registration of the "SBRef" volume using a customized FLIRT BBR algorithm and bbregister for fine-tuning. The intensity modulations of the distortion correction were stored but not applied to the data. The EPI frames were resampled to atlas space using one-step spline resampling, incorporating all necessary transforms such as motion, EPI distortion, EPI to T1w registration, fine-tuning, and nonlinear T1w to MNI registration. Additionally, intensity normalization to a mean of 10000, similar to FEAT, and bias field removal were performed. Finally, a brain mask based on FreeSurfer segmentation was applied.

The cortical ribbon was mapped from MNI space 2mm volume to MNI space native mesh surface, including voxels between the white and pial surface and partially weighted voxels. Voxels with a high temporal coefficient of variance were excluded to remove small vessels and brain rim voxels. The time series was transformed to the fs_LR registered 32k mesh and underwent surface-based and subcortical parcel-constrained smoothing with a 2mm FWHM. A dense time series was created on a standard set of brain ordinates [40].

## 3. RESULTS

In our data analysis from 138 participants, we extracted the connectivity matrix. We applied the sign matrix within the framework of structural balance theory, focusing on positive and negative links, balanced and imbalanced triads, the Tendency to Make Hub (TMH), and total brain network energy. Outliers were removed using the Interquartile Range (IQR) method, and we employed the Wilcoxon test due to the non-normality of the data. The study examined three conditions, namely, resting-state and two emotional processing conditions (fear and neutral), using the Glasser parcellation (360 brain nodes).

Our results unveiled a compelling trend in the number of positive links across the conditions. The number of positive links surged significantly during emotional processing, a finding that piqued our interest. Surprisingly, there was no significant difference between the two emotional conditions. Specifically, the increase from resting-state to neutral condition ($p = 0.0030$) and resting-state to fear condition ($p = 0.0402$) were both statistically significant, while the neutral vs. fear comparison was not ($p = 0.0834$).

We observed a significant decrease during emotional processing compared to the resting-state for negative links, mirroring the pattern seen with positive links. The decrease was significant for both the neutral condition ($p = 0.0030$) and the fear condition ($p = 0.0402$), though the difference between neutral and fear conditions remained non-significant ($p = 0.0834$).

Our investigation into the number of balanced and imbalanced triads revealed an interesting trend. Balanced triads, a marker of brain stability, significantly decreased during emotional processing compared to the resting-state. This decrease was significant for both the neutral ($p = 0.0226$) and fear conditions ($p = 3.10 \times 10^{-5}$). Notably, the difference between the neutral and fear conditions was also significant ($p = 0.0123$). Conversely, the number of imbalanced triads, a potential indicator of instability, increased during emotional processing, with statistically significant increases for both neutral ($p = 0.0226$) and fear conditions ($p = 3.10 \times 10^{-5}$) and a significant difference between the two emotional conditions ($p = 0.0123$).

We also identified regions with a tendency to serve as influential hubs, which are nodes in a network that have a high degree of connectivity with other nodes. During the resting-state, there were more hubs compared to emotional processing conditions, indicating a more distributed network. The number of hubs was significantly lower during fear processing than during the neutral condition, suggesting a more centralized network during fear processing. TMH differences between the resting-state and neutral ($p = 2.59 \times 10^{-8}$), resting-state and fear ($p = 1.43 \times 10^{-11}$), and neutral vs. fear conditions ($p = 0.00173$) were all statistically significant. Similar trends were observed for positive TMH (posTMH), with significant differences between resting and neutral ($p = 9.62 \times 10^{-20}$), resting and fear ($p = 4.16 \times 10^{-20}$), and neutral vs. fear conditions ($p = 0.000179$).

Finally, our examination of overall balance-energy shed light on the intensity of emotional processing. Total balance-energy was lower during the resting-state compared to emotional tasks, with higher energy levels observed during fear processing than neutral. Energy differences between resting and neutral (p = 0.0226), resting and fear (p = 3.10×10^-5), and neutral vs. fear (p = 0.0123) conditions were all statistically significant.

## 4, DISCUSSION

In this study, we indicated that the stability of the brain's functional network altered during emotional processing, and the functional links' topology could influence the network's stability. In this regard, we compared the quality of triadic associations during emotional processing to the condition of the resting state. In addition, our emotional processing has been investigated in two different states: neutral and fear. We showed that the resting state has lower balance energy as compared to the emotional processing states. Moreover, balance energy in the fear condition was significantly higher than in the neutral condition. We also showed that the value of negative TMH is higher in the resting state compared to the emotional processing conditions. Since signed link interactions between different types of triadic have an important role in network stability, we decided to explore the quality of balanced and imbalanced triads in both resting-state and emotional processing conditions.

### 4.1. Dynamic Adaptability and Balance of Brain Networks

Our results underscore the intricate adaptability of brain networks when processing emotional stimuli, highlighting a notable shift between balanced and imbalanced triads across different states. The observed decrease in balanced triads during emotional conditions, whether neutral or fear-induced, suggests that emotional processing disrupts network stability. This finding points to the brain's remarkable capacity to modulate its functional organization in response to varying emotional demands, emphasizing an adaptive mechanism where network stability is flexibly adjusted. This adaptability could be essential for efficiently managing emotional responses, allowing the brain to calibrate its internal structure to accommodate the nuances of different emotional states [41,42].

### 4.2. Role of Tendency to Make Hub (TMH) in Emotional States

Previous studies have suggested that in signed networks, the specific topological arrangement of positive and negative links, particularly the distribution of negative links around hubs, contributes to lower balance-energy and a more stable network state near absolute stability [21]. Our results show that the negative TMH during emotional processing and the fear condition is lower than in the resting state, leading to an increase in total balance energy and a more unstable network. This increase in energy is a result of the brain creating a demand to change its state and ultimately increase its energy. Also, the decreased tendency to form hubs (TMH) during emotional processing, particularly under fear conditions, underscores a shift toward a more centralized network structure. This change suggests a processing style that emphasizes efficiency over widespread connectivity, potentially facilitating a focused response when encountering emotionally charged situations. Such a reconfiguration could support immediate, survival-focused responses, yet also holds implications for understanding how fear-related disorders, like PTSD, may emerge from persistent centralization in the brain's network [29,43].

## 4.3. Effect of imbalanced triads on the balance

Our research findings indicate a significant increase in the level of imbalanced brain quantity. As these imbalanced triads grow, the brain's balance-energy level also increases, resulting in a more unstable functional network of the brain [29]. These findings highlight the crucial role of the organization of signed links in enabling the brain to control and regulate its organizational connections, ultimately influencing the brain's functionality and efficiency [44].

## 4.5. Triad Interactions and Network Flexibility

Finally, our research, in line with [20], suggests that the quantity of imbalanced triad interactions can serve as a benchmark for assessing the need for change within the functional brain network. We have validated this assertion by demonstrating that imbalanced triads play a crucial role in modifying the brain's functional network. This finding has significant implications for understanding the brain's adaptability to varying conditions, which is facilitated by imbalanced associations that lead to network modifications. These modifications, in turn, are considered factors for the brain's flexibility to face emotional stimuli, highlighting the broader implications of our research for the field of neuroscience. These findings illuminate a complex, adaptive mechanism in the brain's functional networks, where balanced and imbalanced triads, energy dynamics, and hub tendencies coalesce to facilitate nuanced responses to emotion. Our study contributes to a broader understanding of emotional processing, opening pathways for future research into emotional resilience, particularly in developing targeted interventions to foster balanced network structures.

**Contributions:**
**Sepehr Gourabi**: designed the study, coded, analyzed the data, visualized the concepts, wrote the manuscript, edited the manuscript
**Parinaz Khosravani**: literature review, wrote the manuscript, edited the manuscript, revised the manuscript
**Shahrzad Nosrat**: literature review, wrote the manuscript, edited the manuscript, revised the manuscript
**Roya Mohammadi**: literature review, wrote the manuscript, edited the manuscript, drew the figures
**Masoud Lotfalipour**: wrote the manuscript

**Data Availability**
The data and analysis code used in this study are available upon request. Interested individuals can contact the author via email at sgourabi@binghamton.edu .

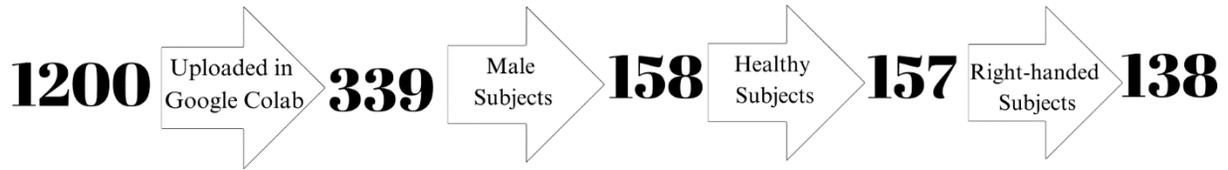

*Figure 1 _ Participant selection process*

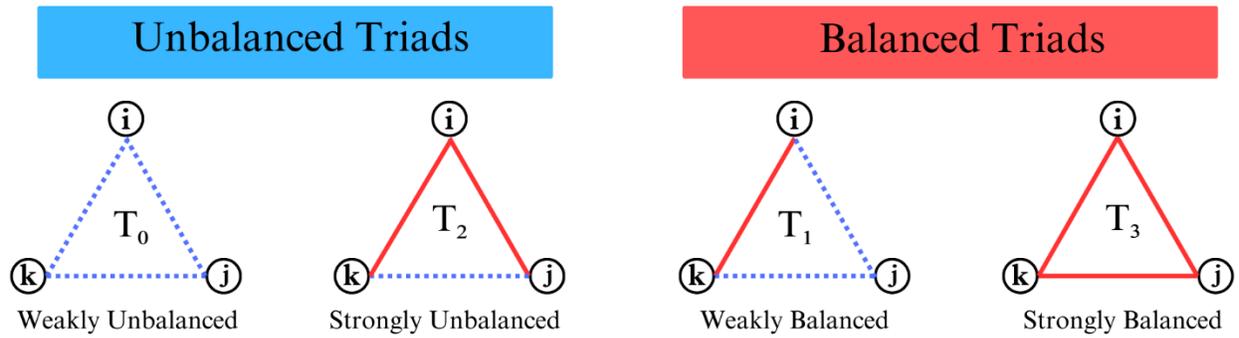

*Figure 2 _ Types of Balanced and Imbalanced Triads*

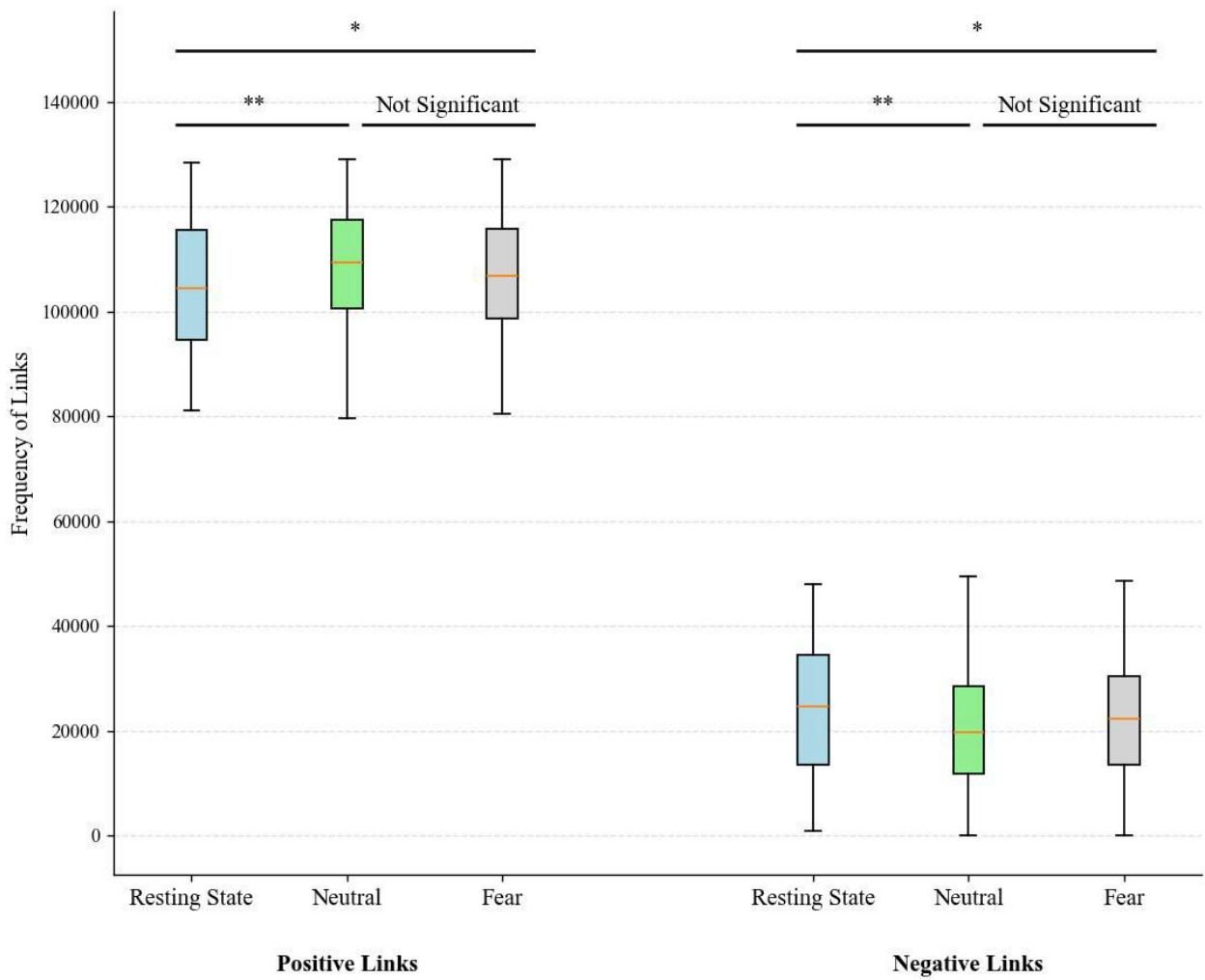

*Figure 3 _ Box plot of positive & negative links*

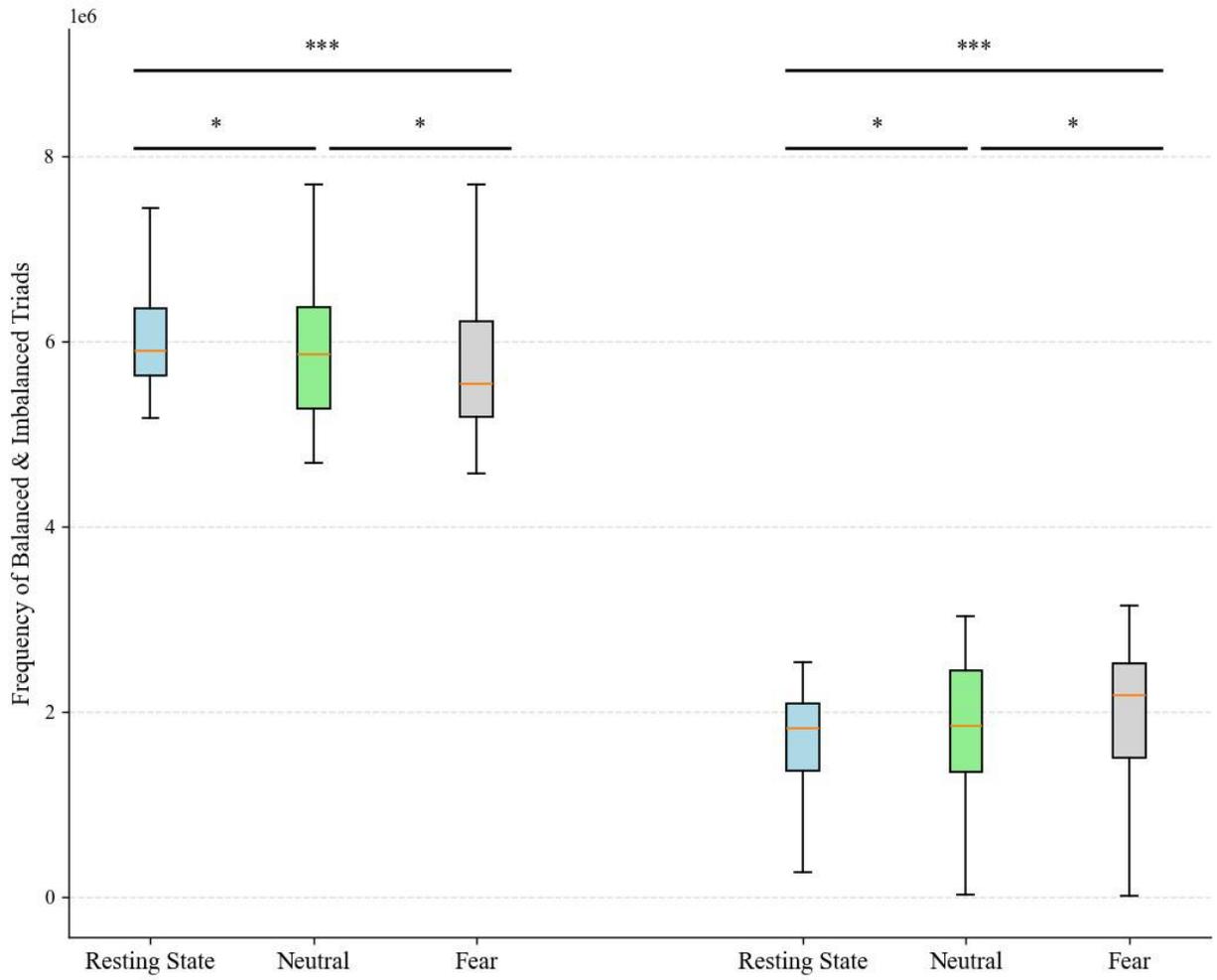

*Figure 4 _ Box plot of Balanced & Imbalanced Triads*

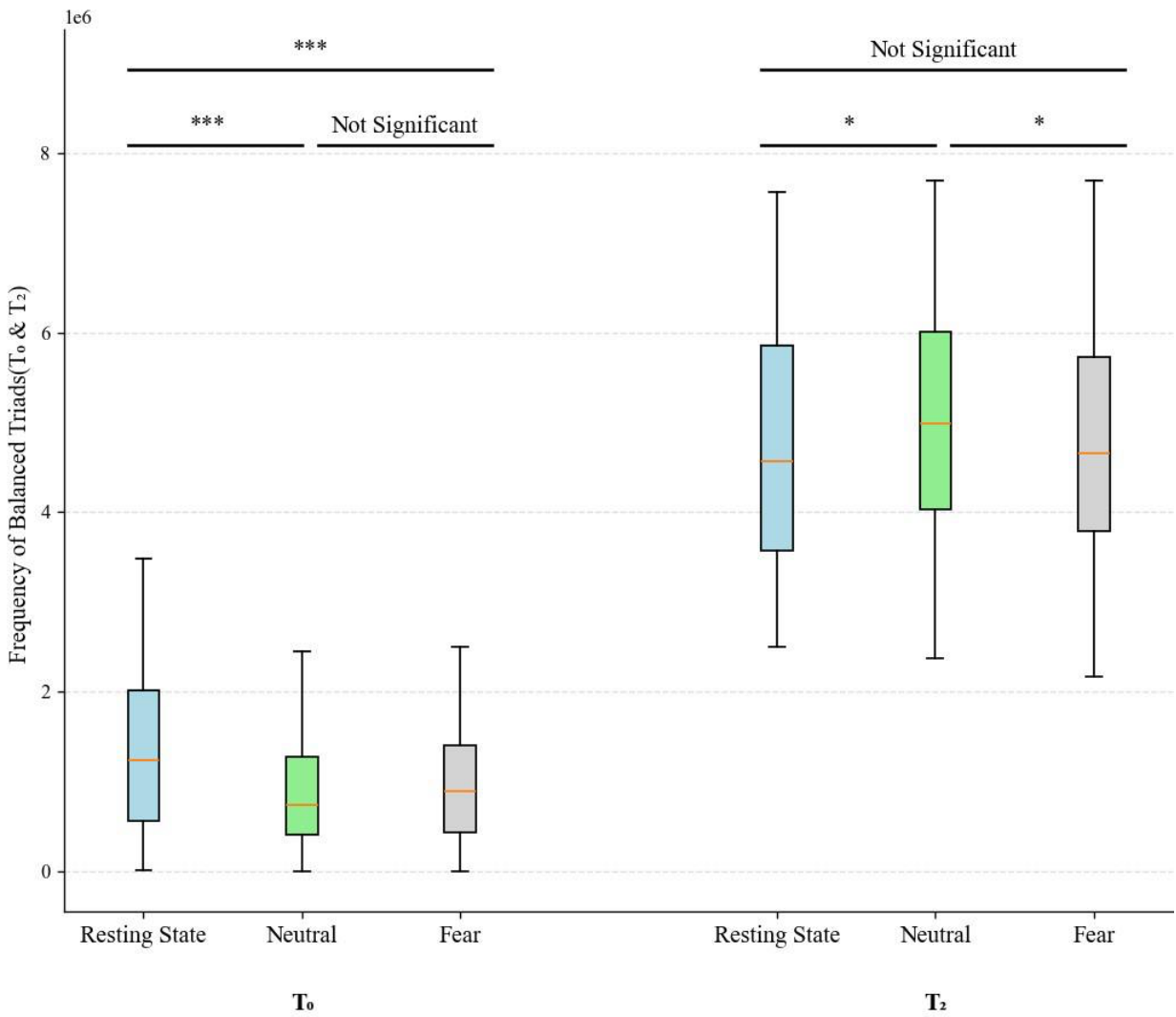

*Figure 5 _ Box plot of different types of Balanced Triads*

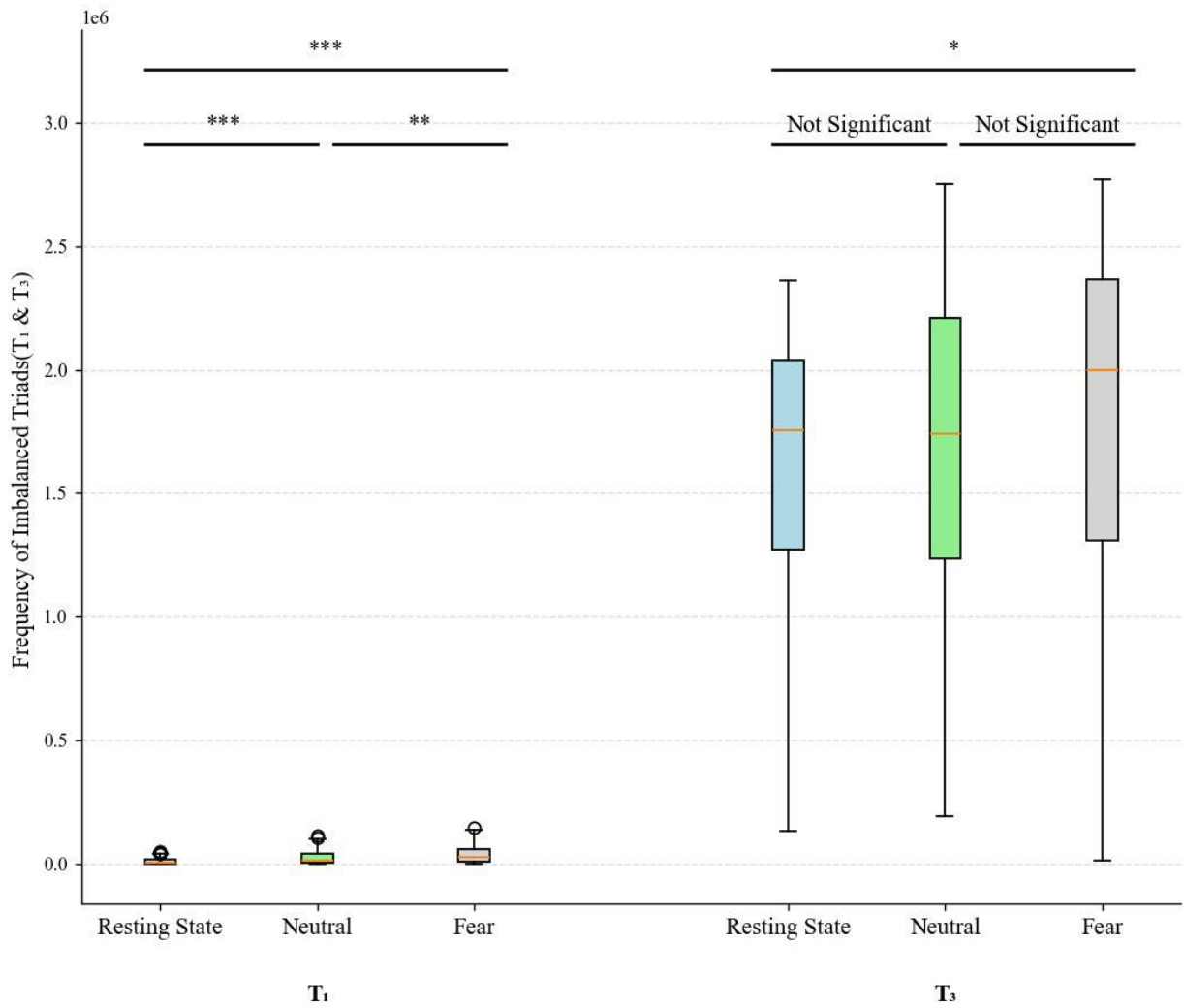

*Figure 6 _ Box plot of different types of Imbalanced triads*

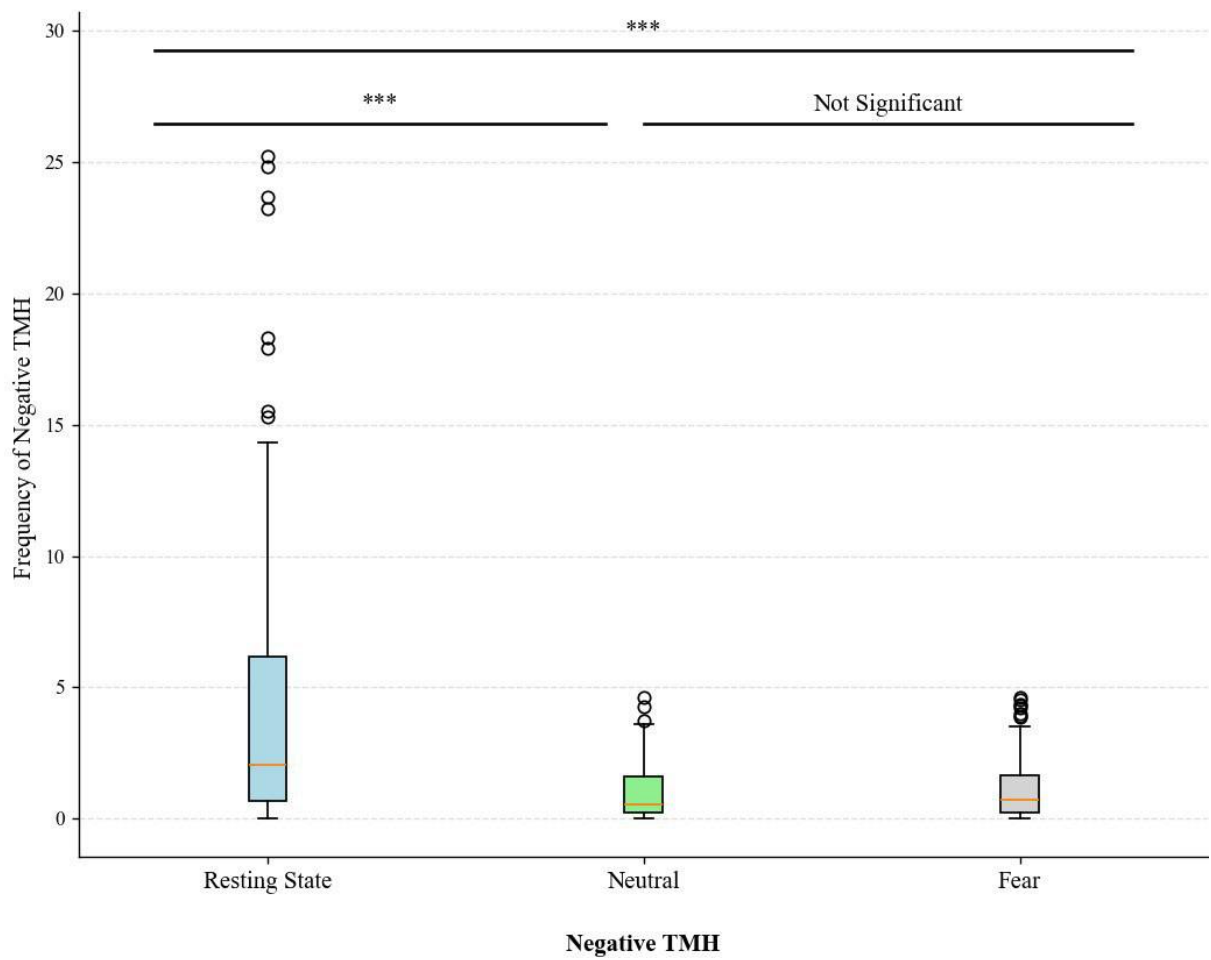

*Figure 7 _ Box plot pf Negative TMH*

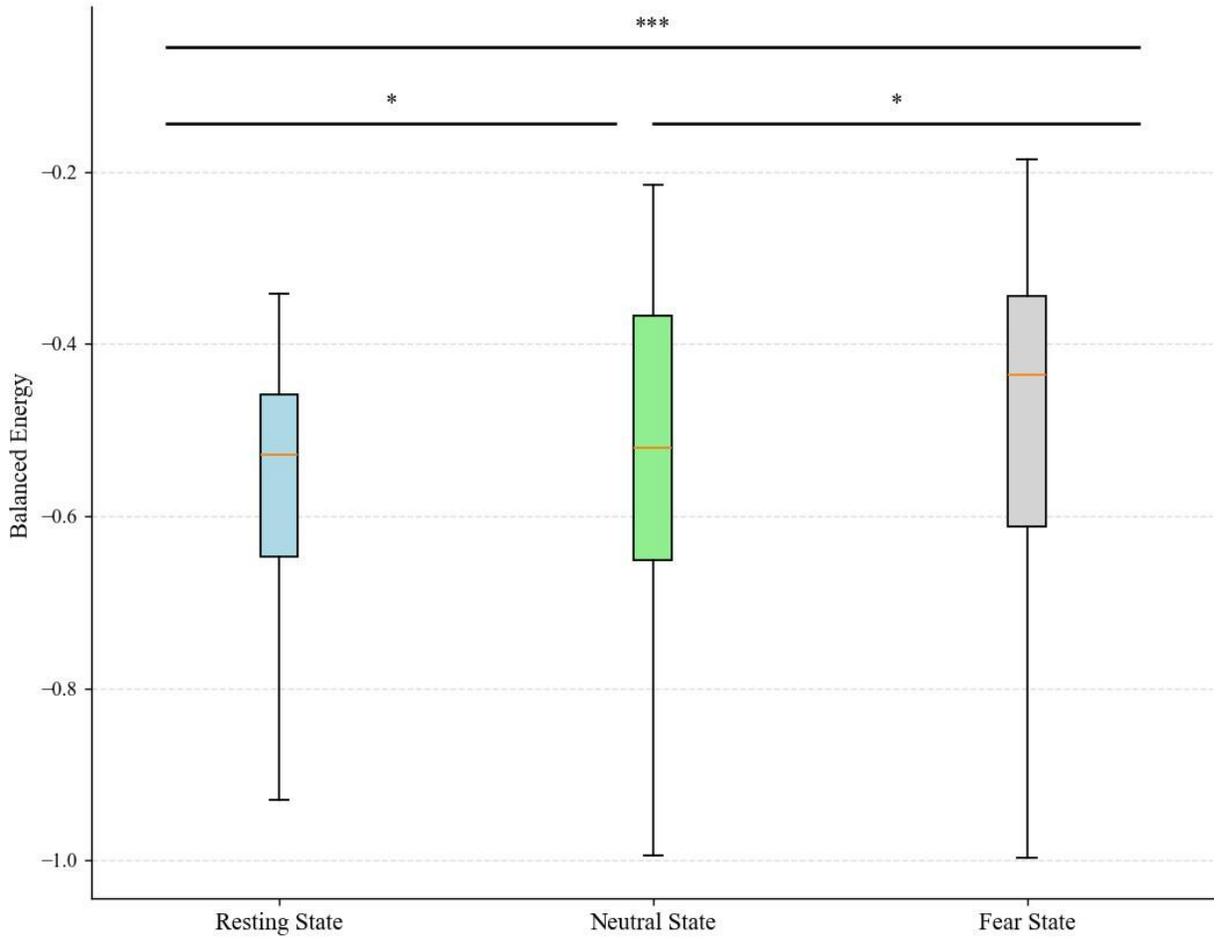

*Figure 8 _ Box plot of Balanced Energy*